\documentstyle[12pt]{article}
\begin{document}

\title{Noncommutative Dynamics of Random Operators}

\author{Michael Heller\thanks{Vatican Observatory, V-00120
Vatican City State. Correspondence address: ul. Powsta\'nc\'ow
Warszawy 13/94, 33-110 Tarn\'ow, Poland. E-mail:
mheller@wsd.tarnow pl} \and Leszek Pysiak\thanks{Faculty of
Mathematics and Information Science, Warsaw University of
Technology, Plac Politechniki 1, 00-661 Warsaw, Poland.}\\
\and Wies{\l}aw Sasin\thanks{Faculty of
Mathematics and Information Science, Warsaw University of
Technology, Plac Politechniki 1, 00-661 Warsaw, Poland.}}

\maketitle

\begin{abstract}
We continue our program of unifying general relativity and
quantum mechanics in terms of a noncommutative algebra ${\cal
A}$ on a transformation groupoid $\Gamma = E \times G$ where $E$
is the total space of a principal fibre bundle over spacetime,
and $G$ a suitable group acting on $\Gamma $. We show that every
$a \in {\cal A}$ defines a random operator, and we study the
dynamics of such operators. In the noncommutative regime, there
is no usual time but, on the strength of the Tomita-Takesaki
theorem, there exists a one-parameter group of automorphisms of
the algebra ${\cal A}$ which can be used to define a state
dependent dynamics; i.e., the pair $({\cal A}, \varphi )$, where
$\varphi $ is a state on ${\cal A}$, is a ``dynamic object''.
Only if certain additional conditions are satisfied, the
Connes-Nikodym-Radon theorem can be applied and the dependence
on $\varphi $ disappears. In these cases, the usual unitary
quantum mechanical evolution is recovered. We also notice that
the same pair $({\cal A}, \varphi )$ defines the so-called free
probability calculus, as developed by Voiculescu and others,
with the state $\varphi $ playing the role of the noncommutative
probability measure. This shows that in the noncommutative
regime dynamics and probability are unified. This also explains
probabilistic properties of the usual quantum mechanics.We continue our program of unifying general relativity and
quantum mechanics in terms of a noncommutative algebra ${\cal
A}$ on a transformation groupoid $\Gamma = E \times G$ where $E$
is the total space of a principal fibre bundle over spacetime,
and $G$ a suitable group acting on $\Gamma $. We show that every
$a \in {\cal A}$ defines a random operator, and we study the
dynamics of such operators. In the noncommutative regime, there
is no usual time but, on the strength of the Tomita-Takesaki
theorem, there exists a one-parameter group of automorphisms of
the algebra ${\cal A}$ which can be used to define a state
dependent dynamics; i.e., the pair $({\cal A}, \varphi )$, where
$\varphi $ is a state on ${\cal A}$, is a ``dynamic object''.
Only if certain additional conditions are satisfied, the
Connes-Nikodym-Radon theorem can be applied and the dependence
on $\varphi $ disappears. In these cases, the usual unitary
quantum mechanical evolution is recovered. We also notice that
the same pair $({\cal A}, \varphi )$ defines the so-called free
probability calculus, as developed by Voiculescu and others,
with the state $\varphi $ playing the role of the noncommutative
probability measure. This shows that in the noncommutative
regime dynamics and probability are unified. This also explains
probabilistic properties of the usual quantum mechanics.
\end{abstract}

\section{INTRODUCTION}
In a series of works (Heller et al., 1997, 2000; Heller and
Sasin 1999) we have formulated a program to unify general
relativity and quantum mechanics based on a noncommutative
algebra on a transformation groupoid.  In (Heller et al. 2004a)
we have tested the program by constructing a simplified (but
still mathematically interesting) model and computing many of
its details, and in (Heller et al. 2004b) we have discussed its
observables with a special emphasis on the position and momentum
observables. In the present work, we study its dynamic and
probabilistic aspects.

Let us first, for the reader's convenience, outline the main
architectonic properties of our model. We construct a
transformation groupoid in the following way. Let $\tilde{E}$ be
a differential manifold and $\tilde{G}$ a group acting smoothly
and freely on $\tilde{E}$.  We thus have the bundle $(\tilde{E},
\pi_M, M = \tilde{E}/\tilde{G})$, and we can think of it as of the
frame bundle, with $\tilde{G}$ the Lorentz group, over spacetime
$M$. To simplify our construction, we choose a finite subgroup
$G$ of $\tilde{G}$ and a cross section $S: M
\rightarrow \tilde{E}$ of the above bundle (it need not be
continuous). Then we define $E = \bigcup_{x \in M}S(x)G$. The
free action of $G$ (to the right) on $E$, defines the
transformation groupoid structure on the Cartesian product
$\Gamma = E \times G$ [for details see (Heller et al. 2004a)].
The choice of the cross section $S: M \rightarrow \tilde{E}$ can
be regarded as the choice of a gauge for our model.

The elements of the groupoid $\Gamma $ represent symmetry
operations of the model. The noncommutative algebra ${\cal A} =
C^{\infty }(\Gamma , {\bf C})$ of smooth complex valued
functions on $\Gamma $ (if necessary, we shall assume that they
vanish at infinity) with convolution as multiplication is an
algebraic counterpart of this symmetry space. In the previous
works, we have reconstructed geometry of the groupoid $\Gamma =
E \times G$ in terms of this algebra. By projecting the full
geometry onto the $E$-direction we recover the usual spacetime
geometry and, consequently, the standard general relativity. The
regular representation $\pi_p: {\cal A} \rightarrow {\rm
End}({\cal H}_p)$ of the groupoid algebra ${\cal A}$ in a
Hilbert space ${\cal H}_p$, for $p\in E$, gives the
$G$-component of the model which can be considered as its
quantum sector.
\par
In the present paper, we show that every $a \in {\cal A}$
defines a random operator (Section 2), and we study the dynamics
of these operators (Section 3). This is not a trivial task.
Noncommutative spaces are nonlocal entities. In general, the
concepts such as that of point and its neighborhood are
meaningless in them.  Therefore, in the noncommutative setting
the concept of the usual ``coordinate time'' is not applicable,
and the question concerning the existence of dynamics arises.
However, as shown by Alain Connes (1994) the algebra ${\cal A}$
admits, on the strength of the Tomita-Takesaki theorem, a
one-parameter group of automorphisms of ${\cal A}$ (called the
{\it modular group\/}), and this group can be used to define a
``modular dynamics'' (Connes and Rovelli, 1994).  But, strangely
enough, this dynamics depends of the state $\varphi $ on the
algebra ${\cal A}$, and only if certain additional conditions
are satisfied, the dependence on $\varphi $ disappears, and one
recovers the usual unitary evolution, well known from quantum
mechanics (Section 4).
\par
In Section 5, we briefly recall the noncommutative probability
calculus (called also {\it free probability calculus\/}) as it
was introduced by Voiculescu (1985), and developed by others
(Voiculescu et all., 1992; Biane, 1998). Such a probability is
defined as the pair $({\cal A}, \varphi )$ where ${\cal A}$ is
an associative algebra (with unity), and $\varphi $ is a state
on ${\cal A}$, i.e., a positive linear functional on ${\cal A}$
such that $\varphi (1) = {\bf 1}$. We can think of $\varphi $ as
of a probability measure on ${\cal M}$. We thus see that the
pair $({\cal A}, \varphi )$ is both the ``dynamic object'' and
the ``probabilistic object''. It follows that, in our model,
every dynamics is probabilistic (in the generalized sense), and
every (generalized) probability has a dynamic aspect. This
important property of the noncommutative regime, supposedly
governing the fundamental level of physics, is inherited by the
quantum sector of our model. In this way, the probabilistic
character of the standard quantum mechanical (unitary) evolution
is explained.
\par
Finally, in Section 6, we briefly comment on the obtained
results.

\section{ALGEBRA OF RANDOM OPERATORS}
Let $\Gamma = E \times G$ be the groupoid described in the
Introduction. In this paper, unless explicitly stated otherwise,
$G$ will always be a finite group. We consider the algebra
${\cal A} = C^{\infty }(\Gamma , {\bf C})$ of smooth complex
valued functions on $\Gamma $ with the convolution as
multiplication. If $a, b \in {\cal A}$, the convolution is
defined as
\[
(a * b)(\gamma) = \sum_{\gamma_1 \in \Gamma_{d{\gamma }}}
a(\gamma \circ \gamma_1^{-1}) g(\gamma_1)
\]
where $\gamma , \gamma_1 \in \Gamma $, and $\Gamma_{d{\gamma}}$
denotes the fiber of the groupoid $\Gamma $ over $d(\gamma) =
d(p, g) = p \in E$ with $g \in G$ [for details see (Heller et
al., 2004a)].
\par
Every $a\in {\cal A}$ generates a random operator $r_a=
(\pi_p(a))_{p\in E}$. It acts on a collection $\{{\cal
H}_{}^p\}_{p\in E}$ of Hilbert spaces $ {\cal
H}^p=\mbox{$L^2$$(\Gamma^p)$}$. Here $\Gamma ^p$ denotes the fiber
of $\Gamma $ consisting of all its elements ``ending at'' $p \in
E$. Every operator $\pi_ p(a)$ is a bounded operator on ${\cal
H}^p$.

An operator $r_a$ to be random must satisfy the following
conditions:

(1)If $\xi_p,\eta_p\in{\cal H}^p$ then the function
$E\rightarrow {\bf C}$ given by
\[E\ni p\mapsto ((r_a)_p\xi_p,\eta_p),\]
for $a\in {\cal A}$, is measurable in the usual sense (i.e.,
with respect to the standard manifold measure on $E$). In our
case this condition is always satisfied.
\par
(2) The operator $r_a$ must be {\em bounded}, i.e., $||
r_a||<\infty$$ $ where
\[||r_a||=\,{\rm e}{\rm s}{\rm s}\,{\rm s}{\rm u}{\rm p}
||\pi_p(a)||.\]
Here ``ess sup'' denotes essential supremum, i.e.,  supremum
modulo zero measure sets. Let us notice that if, in our case,
$a$ is a bounded function, condition (2) is satisfied, and if
$a$ is continuous, condition (1) is satisfied.
\par\
Let ${\cal M}$ be the $*$-algebra of equivalence classes
(modulo equality almost everywhere) of bounded random
operators $(A_p)_{p\in E}$ equipped with the following operations:
\begin{enumerate}\item($A+B)_p=A_ p+B_p,$ \item
$(A^{*}_p)=(A)^{*}_p$, \item$(A\cdot B)_p=A_p\cdot B_p$,
\end{enumerate}$
A,B\in {\cal M},\,p\in E$.  The well known result is that ${\cal
M}$ forms a von Neumann algebra, i.e.,  $ {\cal M}={\cal
M}^{\prime\prime}$ where ${\cal M}^{\prime\prime}$ denotes the
double commutant of $ {\cal M}$ (Connes, 1994, p.  52).  This
result clearly applies to our case, i.e.,  random operators $r_
a$ defined above form a von Neumann algebra.  We will call it
the {\em von Neumann algebra of the groupoid} $\Gamma$.
\par
In the matrix representation we have (Heller et al., 2004a)
\[L^{\infty}(\Gamma ,{\bf C})\simeq L^{\infty}(M)\otimes
M_{n\times n}({\bf C}).\] In this representation, the von
Neumann algebra of random operators assumes the form
\[{\cal M}\simeq L^{\infty}(E)\otimes M_{n\times n}({\bf C}
)\simeq M_{n\times n}(L^{\infty}(E)).\]

Let us now recall some terminology.  Let $\varphi$ be a positive
linear functional on a von Neumann algebra ${\cal M}$: $\varphi$
is said to be {\em faithful\/} if $ 0\neq x\in {\cal M}$ implies
$\varphi (x)>0$; $\varphi $ is said to be {\em normal\/} if $
\varphi (x)={\rm s}{\rm u}{\rm p}\varphi (x_i)$ provided
$x$ is the supremum of a monotonically increasing net $
\{x_i\}$ in the collection of
positive operators in ${\cal M}$; $\varphi$ is called {\em
tracial\/} if $\varphi (x^{*}x)=\varphi (xx^{ *})$ for every
$x\in {\cal M}$; $\varphi$ is said to be a {\em state\/}  if it
is positive and normed to unity.
\par
A von Neumann algebra ${\cal M}$ is called {\em finite\/} if it
admits a faithful, normal and tracial state.

In our case, continuous functionals on ${\cal M}$ are tracial
and are of the following form
\[\varphi (r_a)=\int {\rm T}{\rm r}(r_a(p)\rho (p))d\mu_
E(p)\] for $r_a\in {\cal M}$, where $\rho\in L^1(E)\otimes M_{
n\times n}({\bf C})\simeq M_{n\times n}(L^1(E))$ or,
equivalently, with the dependence on $x\in M$ clearly displayed
\[\varphi (r_a)=\int_M\sum_{g\in G}{\rm T}{\rm r}(r(s(x
)\cdot g)\cdot\rho (s(x)\cdot g))d\mu (x).\]
\par
For $\varphi (r_a)$ to be positive, $\rho (p)$ must be a
positive matrix, i.e., having all its eigenvalues non-negative,
for almost all $ p\in E$. If all eigenvalues of $\rho (p)$ are
positive, the state is faithful.
\par
Let us define the normalization: if $r_a(p)={\bf 1}$, for every
$ p\in E$, then
\[\varphi (r_a)=\int_M\sum_{g\in G}\mbox{${\rm T}{\rm r}
(\rho (s(x)\cdot g))d\mu (x)=1${\em .}}\]
\par
{\bf Proposition.} The von Neumann algebra ${\cal M}$ of the
groupoid $\Gamma$ is finite.
\par
{\bf Proof:} Let us choose $\rho (s(x)\cdot g)=f(x)\cdot {\bf 1}$
where $f\in L^1(M)$, and $f>0$. $\rho$ is clearly positive and
faithful. Then normalization condition reduces to the
following formula
\[\varphi (r_a)=\int_Mnf(x)d\mu (x)=1,\]
where $n$ is the rank of the group $G$.  Therefore, $\varphi$ is
a state. It is also a normal state since every fiber in $\Gamma$
is finite, and the normality is a simple consequence of the
Lebesgue majorized convergence theorem. $\Box$

\section{EVOLUTION OF RANDOM OPERATORS}
Now, we define the Hamiltonian $H_p^{\varphi}={\rm L}{\rm o}
{\rm g}\rho_p^{\varphi}$, and the Tomita-Takesaki theorem gives
us the evolution of random operators dependent on the state
$\varphi$ in terms of the one-parameter group of automorphisms
$\sigma_s^{\varphi}$, called {\em modular group} (Connes, 1994,
pp. 43-44, 496-470)
\begin{equation}\label{TT}\sigma_s^{\varphi}(r_a(p)=e^{
isH_p^{\varphi}}r_a(p)e^{-isH_p^{\varphi}}\end{equation} for
every $p\in E$.
\par
Equation (\ref{TT}) can also be written in the form
\begin{equation}\label{TT2}i\hbar\frac d{ds}|_{s=0}\sigma^{
\varphi}_s(r_a(p))=[r_a(p),H_p^{\varphi_{}}].\end{equation}
This equation describes the state dependent evolution of random
operators with respect to the parameter $s\in {\bf R}$ of the
modular group.
\par
Our aim is now to obtain the state independent evolution by
applying to our case the construction based on the
Connes-Nicodym-Radon theorem (Sunder, 1987, p. 74). Let us first
recall some concepts involved in this construction. Let $ {\rm
A}{\rm u}{\rm t}{\cal M}$ be the group of all automorphisms of an
algebra ${\cal M}$, and $\lambda\in {\rm A}{\rm u}{\rm t} {\cal
M}$. An automorphism $\lambda$ is said to be {\em inner\/} if
there exists an element $u\in {\cal U}$, where ${\cal U}=\{u\in
{\cal M} :uu^{*}=u^{*}u=1\}$ is the unitary group of the algebra
${\cal M}$, such that 
\[\lambda (b)=ubu^{*}\] 
for every
$b\in {\cal M}$. Let ${\rm I}{\rm n}{\rm n} {\cal M}$ denote
the group of inner automorphisms of ${\cal M}$.  We define two
automorphisms $\lambda_1$ and $\lambda_2$ to be {\em inner
equivalent\/} if 
\[\lambda_1(b)=u\lambda_2(b)u^{*},\]
for every $b\in {\cal M}$ and the group ${\rm O}{\rm u}{\rm
t}{\cal M}$ of {\em outer automorphism\/} as
\[{\rm O}{\rm u}{\rm t}{\cal M}:={\rm A}{\rm u}{\rm t}{\cal M}
/{\rm I}{\rm n}{\rm n}{\cal M}.\]
\par
Let $\sigma^{\varphi}_s,\sigma^{\psi}_s\in {\rm A}{\rm u} {\rm
t}{\cal M}$ for a fixed $t\in {\bf R}$, and let us further
assume that there exists the unitary element $u\in {\cal U}$
such that
\[\sigma^{\psi}_s=u\sigma^{\varphi}_su^{*}.\]
Hence,
\[[\sigma^{\psi}_s]=[\sigma^{\varphi}_s]\]
where square brackets denote the equivalence class of a given
automorphism.  If we define the canonical homomorphism
\[\delta :{\bf R}\rightarrow {\rm O}{\rm u}{\rm t}{\cal M}\]
by
\[\delta (s)=[\sigma^{\varphi}_s],\]
we obtain the modular group which is now state independent.
\par
In our case, we clearly have the state dependent evolution as
described by equation (\ref{TT}). Could it be made state
independent by the above procedure? Equation (\ref{TT}) implies
that $\sigma^{\varphi}_s\in {\rm I}{\rm n}{\rm n}{\cal M}$, and
consequently $
\delta (s)={\bf 1}$. This means that the one-parameter
group $\sigma_s$ independent of state is trivial.
\par
This can also be deduced from the Dixmier-Takesaki theorem
(Connes, 1994, p. 470). Let us define
\[S({\cal M})=\{S_0\in {\bf R}:\sigma^{\varphi}_{S_0}\in
{\rm I}{\rm n}{\rm n}{\cal M}\}.\]
The Dixmier-Takesaki theorem says that $S({\cal M})={\bf R}$ if
and only if the algebra ${\cal M}$ is finite (or semifinite, if
the theorem is formulated for weights rather than for states,
see below). And, as we know from the previous section, this is
indeed the case.
\par
The above result means that every $\sigma^{\varphi}_s$ is
unitary equivalent to $ {\rm i}{\rm d}_s$. In other words, the
state independent time does exist, but nothing changes in it.
This fact is clearly the consequence of the oversimplified
character of our model; in particular, of the fact that the
group $G$ is finite.

\section{QUANTUM AND CLASSICAL DYNAMICS}
So far we have shown that on the fundamental (noncommutative)
level we have a state dependent ``modular dynamics'' which (at
least in more realistic models) can be made state independent.
Now, the questions arise: what do we get of this dynamics, if we
go to the quantum sector and the spacetime sector of our model,
respectively?
\par
To answer the first of these questions, let us restrict the von
Neumann algebra ${\cal M}$ to its subalgebra
\[{\cal M}_G=\{f\circ pr_G:f\in {\bf C}^G\}\]
where $pr_G:\Gamma\rightarrow G$ is the obvious projection. For
every $ a\in {\cal M}_G$, the random operator
$r_a=(\pi_p(a))_{p\in M}$ is a family of operators which can be
identified with each other (on the strength of the natural
isomorphism $
\Gamma_p\simeq G$).
Therefore, any such $r_a$ is a family projectible to a single
operator on ${\cal H}_G=L^2(G)$. The operator on ${\cal H}_G$ to
which $ r_a$ projects will be denoted by $a_G$.  Let us notice
that it is not a random operator.
\par
For $a_G\in {\rm E}{\rm n}{\rm d}{\cal H}_G$, equation
(\ref{TT2}) assumes the form
\[i\hbar\frac d{ds}|_{s=0}(\sigma^{\varphi}_s(a_G))=[a_
G,H^{\varphi}].\]
The only difference between this equation and the Heisenberg
equation, well known from quantum mechanics, is that this
equation depends on the state $\varphi$. But even this
difference disappears for more realistic models in which
Connes-Nikodym-Radon theorem gives the state independent modular
evolution. In these cases, the standard quantum mechanical
dynamics is recovered.
\par
Now, let us turn to the question of what do we obtain by going
to the spacetime (macroscopic) sector of our model. In (Heller
et al., 2004a) we have shown that the answer can be obtained by
the averaging procedure. Let us consider the von Neumann algebra
${\cal M}$ in its matrix representation, and let $ M_a$ be a
matrix corresponding to the function $a$. Then by averaging of $
M_a$ we understand
\[\langle M_a(p)\rangle =\frac 1{|G|}{\rm T}{\rm r}M_a\]
where $|G|$ denotes the rank of $G$.
\par
In (Heller et al. 2004b) we have proved that
\[\pi_{pg}(a)=L_g\pi_p(a)L_{g^{-1}}\]
where $L_g$ denotes the left translation by $g\in G$. By
applying the trace operation to both sides of this equality we
obtain
\[{\rm T}{\rm r}(\pi_{pg}(a))={\rm T}{\rm r}(\pi_p(a)),\]
i.e., the averaging operation gives a function on $M$, and
equation (\ref{TT}) reduces to
\[\langle\sigma^{\varphi}_s(r_a(p)\rangle =\langle r_a(
p)\rangle .\]
We see that the dependence on $\varphi $ has disappeared.  This
equation shows that the modular dynamics (with respect to the
parameter $s$) is a quantum phenomenon which is not directly
visible from the spacetime perspective. The ``modular time'' $s$
is related to the usual ``coordinate time'' $ t$ by the
dependence on $p=(x_0,x_1,x_2,x_3,\delta_0,\delta_1,\delta_
2,\delta_3)\in E$.
\par

\section{DYNAMICS AND PROBABILITY}
The fact that dynamics in our model is given in terms of random
operators discloses a link between dynamics and probability.
This link goes much further.
\par\
If $X$ is a compact topological space then there is a strict
correspondence between finite Borel measures on $X$ and linear
forms on the Banach space $C_{}$(X) of continuous functions on
$X$ with the norm:  $\parallel f\parallel ={\rm s}{\rm u} {\rm
p}|f(x)|$, $f\in C_{}(X),\,x\in X$.  Instead of considering the
measure space $(X,m)$, where $m$ is a finite Borel measure on $
X$, we can, equivalently, consider the Banach algebra $C_{}(X)$
together with a distinguished linear form $\varphi$ on it, i.e.,
the pair $ (C_{}(X),\varphi )$.  If, additionally, we impose on
$\varphi$ a suitable normalization condition, this pair will be
a functional counterpart of the probability space.  This is the
starting point of a generalization to the noncommutative concept
of probability.  In place of the Banach algebra $C_{}(X)$ we
consider any associative, not necessarily commutative, unital
algebra ${\cal A}$.  For generality reasons we assume that it is
a complex valued algebra.  Let further $\varphi$ be a linear
(complex valued) form on ${\cal A}$.  If $ $it is a
noncommutative algebra, the pair $ ({\cal A},\varphi )$ is
called the {\em noncommutative probability space}.
Noncommutative probability is also called {\em free
probability\/} (Voiculescu et al. 1992; Biane, 1998).
\par\
However, the above formulated noncommutative probability is too
general for practical purposes.  Some additional conditions are
required.  Also at this stage motivations come from the
commutative case.  Let ${\cal H}$ be a separable Hilbert space,
and $ T$ a bounded self-adjoint operator on ${\cal H}$.  It can
be shown that
\begin{enumerate}
\item There exists the unique (up to equivalence) measure on
the interval $I=[-\parallel T\parallel ,$$\parallel T\parallel
]$ such that
\[f(T)=0\;\;\Leftrightarrow\;\;\int |f|d\mu =0,\]
\item
The algebra $M$ of operators on ${\cal H}$ having the form $
f(T)$, for some bounded Borel function $f$ is a von Neumann
algebra (generated by $T$).
\end{enumerate}

$M$ is a commutative von Neumann algebra naturally isomorphic to
the algebra $L^{\infty}(I,\mu )$ of bounded measurable functions
(modulo almost everywhere) on the interval $I$.
\par
In the view of the above, it is natural to regard the theory of
von Neumann algebras as a noncommutative counterpart of the
measure theory, and to agree for the following definition.  The
{\em noncommutative probability space\/} is a pair $({\cal M}
,\varphi )$  where ${\cal M}$ is a von Neumann
algebra and $\varphi$ a faithful and normal state on ${\cal M}$
(Biane, 1998, Sec. 4). In contrast to the commutative case in
which there is only one interesting measure (the Lebesgue
measure), the noncommutative case exhibits the great richness
and complexity.
\par
As we have seen in Section 3, the pair $( {\cal M},\varphi )$ is
a {\em dynamical object\/} in the sense that it determines the
modular evolution dependent on the state $
\varphi$. But if $\varphi$
satisfies certain natural (and, in general, easy to satisfy)
conditions, the same pair is a ``probabilistic object''.
Therefore, every such dynamics is probabilistic, and every such
probabilistic space has a dynamic aspect. Two structures, which
in the standard mathematics were independent of each other, now
are unified. The state $\varphi$ plays now the role of the
probability measure. It is a remarkable fact that it also
determines the dynamical regime. If we change from the
probability measure $\varphi$ to the probability measure $\psi$,
then we automatically go from the dynamic regime
$\sigma^{\varphi}_ s$ to the dynamic regime $\sigma^{\psi}_s$.
Only when we succeed in obtaining the nontrivial, state
independent evolution $\delta :{\bf R}\rightarrow {\rm O}{\rm
u}{\rm t}{\cal M}$, we get the unique unitary probabilistic
dynamics typical for quantum mechanics (as described by the
Heisenberg equation). Let us notice, however, that the state
$\varphi $ can be interpreted as an expectation value.
Therefore, for two (state dependent) inner equivalent modular
evolutions this expectation value is the same, i.e., state
independent (at least for tracial states).

\section{COMMENTS}
It is interesting to notice that both dynamics and probability
are, from the very beginning, strictly connected with unitarity.
Both dynamics and probability are implemented by a von Neumann
algebra which can be defined as an algebra of operators on a
Hilbert space that are invariant with respect to the group of
unitary transformations. This beautifully harmonizes with the
fact well known from the standard quantum mechanics that
unitarity is closely related to the probabilistic evolution of
quantum systems.
\par
To the physicist it might seem astonishing that the modular
evolution is originally dependent on a state of the considered
system. In fact, it was Carlo Rovelli who proposed a quantum
mechanical model with a state dependent time (Rovelli, 1993) and,
together with Alain Connes, tried to extend this concept to
generally covariant theories, by making the time flow depending on
the thermal state of the system (Connes and Rovelli, 1994). Our
approach is more radical. We closely follow the conclusions of the
Tomita-Takesaki theorem, and assume that on the fundamental level
of physics dynamics is indeed state dependent, and only when we
move to lower energy levels, the von Neumann algebra ${\cal
M}$ becomes more ``coarse'' (in the sense that Aut${\cal
M}$ can be replaced by Out${\cal M}$), the Connes-Nicodym-Radon
theorem can be applied and time independent dynamics emerges.
\par
In our model this evolution is trivial but, as we have seen,
this follows from the simplified character of the model.  The
Sunder's theorem (Sunder, 1987, p. 88) gives us even more
information on the nonexistence of state independent change in
our model. The theorem is formulated for weights rather than for
states, but it {\em a fortiori\/} applies to states. A von
Neumann algebra is said to be {\em semidefinite\/} if it admits
a faithful, normal and tracial weight [for definitions see
(Sunder, 1987, p. 52)].  Roughly speaking a weight $\varphi$ on
${\cal M}$ is {\em semidefinite\/} if there are sufficiently
many elements of ${\cal M}$ at which $\varphi$ assumes a finite
value. The theorem asserts that the following conditions for a
von Neumann algebra $ {\cal M}$ are equivalent:
\par
(i) ${\cal M}$ is semifinite;
\par
(ii) $\sigma^{\varphi}$$_t$ is inner for some faithful, normal
and semifinal weight $
\varphi$
on ${\cal M}$;
\par
(iii) $\sigma^{\varphi}_t$ is inner for every such weight.
\par
In order to have a nontrivial state independent evolution the
von Neumann algebra ${\cal M}$ cannot be semifinite. This can be
obtained by using in our model a locally compact non-unimodular
group such as, for example the Poincar\'e group. Another
possibility would be to employ a noncompact group $G$. In such a
case, one cannot integrate along $G$ to reduce the density $\rho
(s(x)\cdot g)$ to the form $f(x)\cdot {\bf 1}$, and the algebra
$ {\cal M}$ could not be semifinite, even if $G$ is a unimodular
group.
\par
Our model has disclosed quite unexpected connection between
noncommutative dynamics and noncommutative probability. The pair
$({\cal M}, \varphi )$ is both the ``dynamic object'' and the
``probabilistic object''. This fact throws some light onto the
``strange'' dependence of the dynamics of random operators on
the state $\varphi $. The state $\varphi $ is also a probability
measure: if we switch to another state, we switch to another
probability measure, and it seems rather natural that together
with the change of the probability measure, the dynamical regime
of random operators changes as well. Two concepts --- dynamics
and probability --- that are separate in the usual
circumstances, in the noncommutative domain turn out to be but
two aspects of the same mathematical structure.

%\newpage
\vspace{0.3cm}
\noindent
{\bf REFERENCES}
\vspace{0.2cm}

Biane, Ph. (1998). ``Free Probability for Probabilists'', arXiv:
math.PR/9809193

Connes, A. (1994). {\em Noncommutative Geometry,\/} Academic
Press, New York -- London.

Connes, A. and Rovelli, C. (1994). {\it Classical and Quantum
Gravity\/} {\bf 11}, 2899.

Heller, M. and Sasin, W. (1999). {\it Int. J. Theor.
Phys.\/} {\bf 38}, 1619.

Heller, M. Sasin, W. and Lambert, D. (1997). {\it J. Math.
Phys.\/} {\bf 38}, 5840.

Heller, M., Sasin, W. and Odrzyg\'o\'zd\'z, Z. (2000). {\it J.
Math. Phys.\/} {\bf 41}, 5168.

Heller, M., Odrzyg\'o\'zd\'z, Z., Pysiak, L. and Sasin, W.
(2004a). {\it Gen. Rel. Grav.\/} {\bf 36}, 111.

Heller, M., Odrzyg\'o\'zd\'z, Z., Pysiak, L. and Sasin, W. (2004b).
In preparation.

Rovelli, C. (1993). Class. Quantum Grav. {\bf 10}, 1549.

Sunder, V.S. (1987). {\em An Invitation to von Neumann Algebras},
Springer, New York -- Berlin -- Heidelberg.

Voiculecsu, D., V. (1985). In: {\it Operator Algebras and Their
Connections with Topology and Ergodic Theory\/}, Lecture Notes
in Mathematics, vol. 1132, Springer, Berlin -- Heidelberg.

Voiculescu, D.V. Dykema, K. and Nica, A. (1992). {\em Free Random
Variables}, CRM Monograph Series No 1. Am. Math. Soc.,
Providence.

\end{document}